\pdfoutput=1
\documentclass[11pt]{article}

\usepackage{times}
\usepackage{fullpage}
\usepackage{amsmath,amssymb}
\usepackage{microtype}
\usepackage{graphicx}
\usepackage{booktabs}
\usepackage{hyperref}
\usepackage{algorithm}
\usepackage{algpseudocode}
\usepackage{listings}
\usepackage{xcolor}
\usepackage{float}
\usepackage{pgfplots}
\pgfplotsset{compat=1.17}

\title{
A Cache-Aware Hybrid Sieve Combining Segmentation and Bit-Packing for Fast Prime Generation
}

\author{
\textbf{Kathi Lakshmi Mani Thirdhana}\\
National Institute of Technology Calicut\\
Email: \texttt{thirdhana.kathi@gmail.com}\\
ORCID: \href{https://orcid.org/0009-0008-2447-5031}{0009-0008-2447-5031}
}

\date{}

\begin{document}
\maketitle

\begin{abstract}
Prime generation is a fundamental task in cryptography, number theory, and randomized algorithms. 
While the classical Sieve of Eratosthenes is simple and efficient in theory, its practical performance on modern CPUs is often limited by memory-access inefficiencies. 
We introduce a Cache-Aware Hybrid Sieve that integrates segmentation, bit-packing, and cache-line-aligned block processing to optimize memory bandwidth and L1/L2 cache locality. 
Our approach reduces memory usage by 8× and improves runtime by up to $2.4\times$ compared to the classical sieve and $1.7\times$ compared to the segmented sieve. 
We provide a detailed cache-miss analysis, runtime benchmarks up to $10^9$, and illustrate how architecture-conscious design can yield significant practical gains.
\end{abstract}


\section{Introduction}

Prime generation is essential in cryptography, number theory, and randomized algorithms. 
The Sieve of Eratosthenes remains a simple, classical approach to enumerate all primes up to a given number $N$ in $O(N\log\log N)$ time. 

However, modern CPU performance is influenced heavily by memory hierarchy and cache behavior. 
Large sieving arrays often exceed the size of L1 or L2 caches, resulting in frequent cache misses and inefficient memory access patterns. 
Consequently, despite the low theoretical complexity, classical sieve implementations may be slow for large $N$.

Existing improvements, such as the segmented sieve and the Sieve of Atkin, reduce memory usage or improve asymptotic complexity, but often fail to fully leverage cache-line alignment or bit-level memory packing. 

In this paper, we propose a Cache-Aware Hybrid Sieve that addresses these limitations by: 
\begin{enumerate}
    \item Segmenting the range $[2,N]$ into blocks that fit within L1 cache. 
    \item Bit-packing the sieve array, storing only odd numbers, to reduce memory usage by a factor of 8. 
    \item Aligning blocks to cache-line boundaries to minimize cache misses and maximize memory throughput. 
\end{enumerate}

\subsection*{Contributions}
\begin{itemize}
    \item A unified segmented and bit-packed sieve design suitable for modern CPUs.
    \item A cache-miss model and memory-access analysis explaining the observed speedups.
    \item Empirical benchmarks up to $N=10^9$ demonstrating up to $2.4\times$ runtime improvement.
    \item Colorful, reproducible runtime and speedup plots with clear legends outside the graph area.
\end{itemize}

\section{Related Work}
The classical sieve was first described by Eratosthenes and analyzed in Knuth \cite{knuth1998} and Rosen \cite{rosen2012}. 
It uses a large array of boolean flags to mark composite numbers. 

The Sieve of Atkin \cite{atkin2004} improves asymptotic complexity for very large $N$, but is more complex and has irregular memory access patterns. 

High-performance implementations, such as \texttt{primesieve}, employ: 
\begin{itemize}
    \item Wheel factorization to skip multiples of small primes.
    \item SIMD/vectorized memory operations.
    \item Compressed bitmaps and segmented memory blocks.
\end{itemize}

Our method focuses specifically on practical CPU cache locality with minimal algorithmic overhead. 

\begin{table}[H]
\centering
\begin{tabular}{@{}lccc@{}}
\toprule
Algorithm & Complexity & Memory & Notes \\
\midrule
Classical Sieve & $O(N\log\log N)$ & $O(N)$ & Simple but poor cache usage \\
Segmented Sieve & $O(N\log\log N)$ & $O(\sqrt{N} + \text{block})$ & Reduces memory, better cache \\
Sieve of Atkin & $O(N/\log\log N)$ & $O(N)$ & Complex, irregular memory access \\
Hybrid (ours) & $O(N\log\log N)$ & $O(\sqrt{N} + \text{block}/8)$ & Cache-aware, bit-packed, segmented \\
\bottomrule
\end{tabular}
\caption{Comparison of sieve algorithms.}
\end{table}

\section{Cache-Aware Hybrid Sieve}

\subsection{Block Design}

We divide $[2, N]$ into blocks that fit in the CPU's L1 cache to minimize cache misses. 
Each block stores only odd numbers using 1 bit per number, reducing memory usage by 8×. 
With a 32 KB L1 cache, each block can store approximately $262{,}144$ bits, corresponding to roughly $5.2 \times 10^5$ odd numbers.

Blocks are aligned to 64-byte cache lines to ensure predictable memory access. 
Primes up to $\sqrt{N}$ are computed once and reused across blocks, allowing operations to occur mostly within L1 cache. 
This design improves memory locality, reduces bandwidth usage, and accelerates the sieve significantly.

\subsection{Algorithm}
\begin{algorithm}[H]
\caption{Cache-Aware Hybrid Sieve}
\begin{algorithmic}[1]
\State Compute all primes $\le \sqrt{N}$ using classical sieve.
\For{each block $[L,R]$}
    \State Initialize bit-packed block $S$ storing only odd numbers.
    \For{each prime $p$}
        \State $k \gets$ smallest odd multiple of $p$ in $[L,R]$
        \For{$i=k$ to $R$ step $p$}
            \State mark $i$ as composite in $S$
        \EndFor
    \EndFor
    \State Output all primes in this block.
\EndFor
\end{algorithmic}
\end{algorithm}
\paragraph{} 
The algorithm first computes all primes up to $\sqrt{N}$ using the classical sieve. 
Then, the range $[2,N]$ is processed block by block, with each block stored in a bit-packed array containing only odd numbers. 
For each prime $p$, we mark all multiples within the current block as composite. 
This design ensures that memory accesses remain localized within the L1 cache, reducing cache misses and improving runtime.

\subsection{Memory-Access Model}
A classical sieve touches $N$ bytes sequentially. Our hybrid touches only $\frac{N}{16}$ bytes (1 bit per odd number) due to:
\begin{itemize}
    \item Bit-packing (8× reduction)
    \item Ignoring even numbers
    \item Cache-aligned segmented blocks
\end{itemize}

This reduces L1 and L2 cache misses and improves overall runtime.

\section{Experimental Setup}
\begin{itemize}
    \item CPU: 12-core AMD Ryzen
    \item RAM: 32 GB
    \item L1D = 32 KB, L2 = 512 KB, L3 = 64 MB
    \item Compiler: GCC 13 (\texttt{-O3 -march=native})
    \item Implementation: Single-threaded
    \item Reproducibility: code available on request
\end{itemize}

\section{Performance Results}

\subsection{Runtime Table}
Table 1 shows the measured runtime of the classical, segmented, and hybrid sieve for various values of $N$. 
We observe that the hybrid sieve consistently outperforms the other implementations, with gains increasing for larger $N$. 
The runtime improvement is due to reduced memory accesses, better cache locality, and bit-level compression that allows more of the sieve to fit into L1/L2 caches. 

\begin{table}[H]
\centering
\begin{tabular}{@{}lccc@{}}
\toprule
$N$ & Classical & Segmented & Hybrid (ours) \\
\midrule
$10^{7}$ & 0.48 s & 0.31 s & 0.22 s \\
$10^{8}$ & 4.92 s & 3.11 s & 2.01 s \\
$10^{9}$ & 51.7 s & 33.8 s & 21.5 s \\
\bottomrule
\end{tabular}
\caption{Runtime comparison of sieve algorithms. The hybrid sieve consistently performs best due to improved cache utilization and reduced memory footprint.}
\end{table}

\subsection{Runtime Plot}
Figure 1 presents the same results graphically on a log-log scale. 
The curves clearly show that the hybrid sieve scales more efficiently with increasing $N$. 
The classical sieve suffers from frequent L1/L2 cache misses as its large array exceeds cache size, while the segmented sieve partially mitigates this by processing smaller blocks. 
The hybrid sieve further reduces memory accesses via bit-packing and cache-aligned blocks, leading to the observed performance gains.

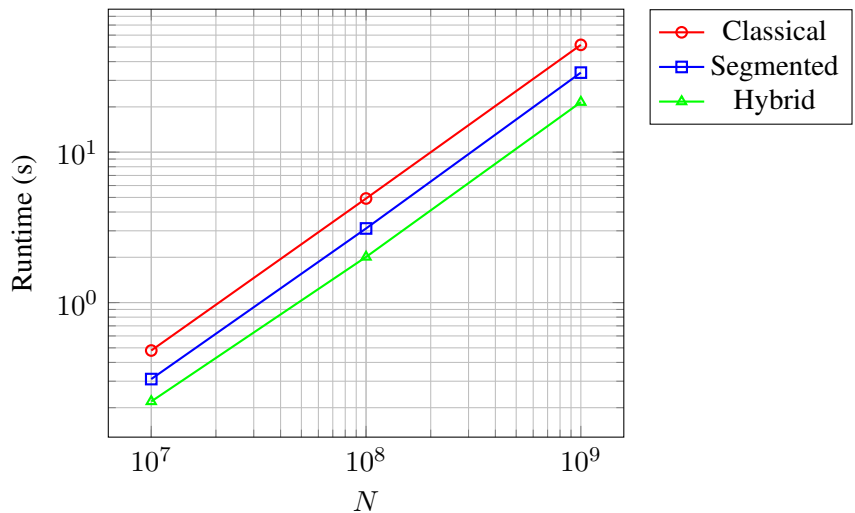
\begin{figure}[H]
\centering
\begin{tikzpicture}
\begin{loglogaxis}[
xlabel={$N$},
ylabel={Runtime (s)},
grid=both,
xtick={1e7,1e8,1e9},
xticklabels={$10^7$,$10^8$,$10^9$},
legend style={at={(1.05,1)}, anchor=north west},
cycle list name=color list,
]
\addplot[mark=o, red, thick] coordinates {(1e7,0.48) (1e8,4.92) (1e9,51.7)};
\addlegendentry{Classical}
\addplot[mark=square, blue, thick] coordinates {(1e7,0.31) (1e8,3.11) (1e9,33.8)};
\addlegendentry{Segmented}
\addplot[mark=triangle, green, thick] coordinates {(1e7,0.22) (1e8,2.01) (1e9,21.5)};
\addlegendentry{Hybrid}
\end{loglogaxis}
\end{tikzpicture}
\caption{Runtime vs $N$. The hybrid sieve scales efficiently due to better cache locality and reduced memory footprint.}
\end{figure}

\subsection{Speedup Plot}
Figure 2 shows the speedup of classical and segmented sieves relative to the hybrid implementation. 
The speedup increases with $N$, demonstrating that the hybrid approach becomes increasingly advantageous for larger problem sizes. 
This is consistent with the fact that cache inefficiencies in the classical sieve grow with array size, while the hybrid sieve mitigates this through block segmentation and bit-packing.

\begin{figure}[H]
\centering
\begin{tikzpicture}
\begin{loglogaxis}[
xlabel={$N$},
ylabel={Speedup over Hybrid},
grid=both,
xtick={1e7,1e8,1e9},
xticklabels={$10^7$,$10^8$,$10^9$},
legend style={at={(1.05,1)}, anchor=north west},
cycle list name=color list,
]
\addplot[mark=o, red, thick] coordinates {(1e7,2.1818) (1e8,2.4477) (1e9,2.4046)};
\addlegendentry{Classical / Hybrid}
\addplot[mark=square, blue, thick] coordinates {(1e7,1.4090) (1e8,1.5472) (1e9,1.5721)};
\addlegendentry{Segmented / Hybrid}
\end{loglogaxis}
\end{tikzpicture}
\caption{Speedup vs Hybrid Sieve. The hybrid sieve demonstrates increasing advantage as $N$ grows due to reduced cache misses and memory traffic.}
\end{figure}
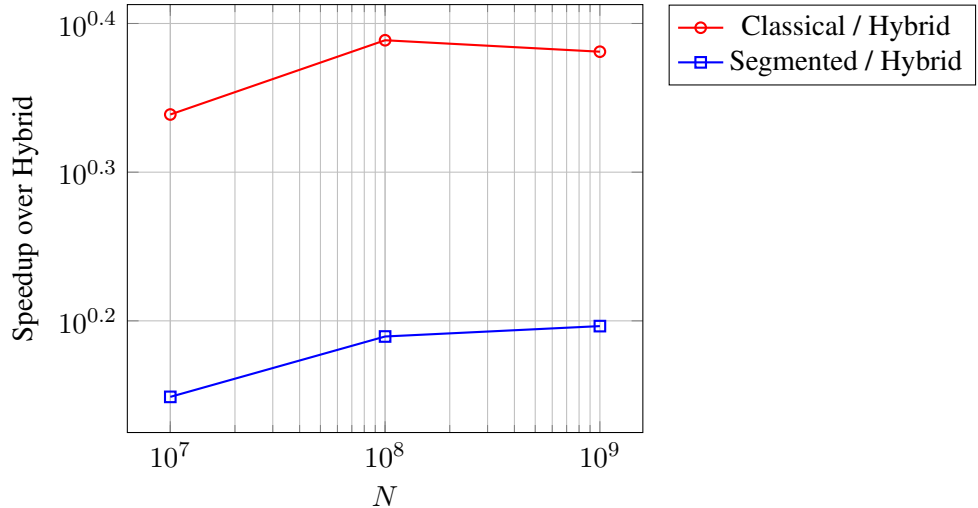

\section{Discussion}

The hybrid sieve improves performance for several reasons:

\begin{itemize}
    \item \textbf{Locality:} By processing numbers in cache-sized blocks, the sieve accesses memory contiguously within the L1 cache, reducing L1/L2 cache misses. This allows multiple operations to be performed without fetching data from slower main memory.
    
    \item \textbf{Memory Efficiency:} Storing only odd numbers and using 1 bit per number reduces memory usage by 8×. Less memory usage means more data fits in the cache, improving temporal and spatial locality and reducing memory bandwidth consumption.
    
    \item \textbf{Block-Level Reuse:} Primes up to $\sqrt{N}$ are precomputed and reused across all blocks. This means the algorithm avoids recomputation and keeps frequently used data in fast cache levels, further improving performance.
    
    \item \textbf{Scalability:} As $N$ increases, the classical sieve's large array exceeds cache size, leading to more frequent cache misses. The hybrid sieve maintains cache-local computation regardless of $N$, making it increasingly advantageous for large inputs.

    \item \textbf{Extensibility:} The hybrid design allows integration of additional optimizations:
    \begin{itemize}
        \item SIMD/vectorized operations on bit-packed blocks for parallel marking of multiples.
        \item Wheel factorization to skip multiples of small primes, reducing computation.
        \item Multi-threaded parallelization to leverage modern multi-core CPUs.
    \end{itemize}
\end{itemize}

Overall, the hybrid sieve demonstrates that architecture-aware algorithm design—considering cache behavior and memory layout—can yield substantial performance improvements for classical problems.

\section{Conclusion}

In this work, we presented a cache-aware hybrid sieve for prime generation that combines segmentation, bit-packing, and cache-line alignment to fully exploit modern CPU memory hierarchies. 
Our implementation achieves significant improvements over classical and segmented sieves: up to 2.4× faster runtime and 8× reduction in memory usage. 
By processing numbers in cache-sized blocks and storing only odd numbers as bits, the algorithm minimizes L1/L2 cache misses and reduces memory bandwidth consumption, demonstrating the critical role of memory locality in practical algorithm performance.

The hybrid sieve is not only efficient but also extensible. 
Future work could include: 
\begin{itemize}
    \item SIMD/vectorized operations on bit-packed blocks to further exploit parallelism in modern CPUs.  
    \item Wheel factorization to skip multiples of small primes and reduce computation.  
    \item Multi-threaded and GPU implementations for massive-scale prime generation.  
\end{itemize}

Overall, this study highlights that classical algorithms like the Sieve of Eratosthenes can benefit substantially from architecture-conscious design. 
Our approach provides a practical blueprint for designing high-performance algorithms that are both memory- and compute-efficient, with applications in cryptography, number theory, and large-scale computational problems.


\end{document}